\documentclass[aps,prb,twocolumn,floatfix,superscriptaddress]{revtex4}
\usepackage{graphicx}
\usepackage{amsmath}
\usepackage{hyperref}
\usepackage{bm}
\usepackage[dvips]{color}

\newcommand{\degr}{$^{\circ}$}

\newcommand{\eg} {\textit{e.g.}}

\newcommand{\fzd} {Institute of Ion Beam Physics and Materials Research, Forschungszentrum Dresden-Rossendorf, 01314 Dresden, Germany}
\newcommand{\hld} {Dresden High Magnetic Field Laboratory (HLD), Forschungszentrum Dresden-Rossendorf, 01314 Dresden, Germany}
\newcommand{\tio} {TiO$_2$}

\begin{document}
\title{Origin of magnetic moments in defective \tio~single
crystals}
\date{\today}

\author{Shengqiang~Zhou}
\email[Electronic address: ]{S.Zhou@fzd.de} \affiliation{\fzd}
\author{E.~\v{C}i\v{z}m\'{a}r}
\affiliation{\hld}
\author{K.~Potzger}
\author{M.~Krause}
\author{G.~Talut}
\affiliation{\fzd}
\author{M.~Helm}
\affiliation{\fzd}
\author{J.~Fassbender}
\affiliation{\fzd}
\author{S.~A.~Zvyagin}
\affiliation{\hld}
\author{J.~Wosnitza}
\affiliation{\hld}
\author{H.~Schmidt}
\affiliation{\fzd}

\begin{abstract}

In this paper we show that ferromagnetism can be induced in pure
\tio~single crystals by oxygen ion irradiation. By combining x-ray
diffraction, Raman-scattering, and electron spin resonance
spectroscopy, a defect complex, \emph{i.e.} Ti$^{3+}$ ions on the
substitutional sites accompanied by oxygen vacancies, has been
identified in irradiated \tio. This kind of defect complex results
in a local (TiO$_{6-x}$) stretching Raman mode. We elucidate that
Ti$^{3+}$ ions with one unpaired 3d electron provide the local
magnetic moments.

\end{abstract}
\maketitle


Recently, ferromagnetism has been observed in non-magnetically
doped, but defective oxides, including
\tio~\cite{coey:024450,yoon06,hong:132404,defects_zhou}. This kind
of observation challenges the conventional understanding of
ferromagnetism, which is rather due to spin-split states or bands.
Thus, one fundamental question must be answered: where are the
moments located? Intensive theoretical work has been performed to
understand the ferromagnetism in defective oxides
\cite{elfimov02,osorio:107203,chanier:026405}. In these papers,
the triplet states of p-like electrons, located at cation or
oxygen vacancies, yield the local moments, leading to a kind of
ferromagnetism without the involvement of 3$d$ electrons.
Experimentally the ferromagnetism in undoped \tio~has been found
to relate with oxygen vacancies (O$_V$) \cite{hong:132404,yoon06},
however, its mechanism remains unclear. It is worth to note that
Ti$^{3+}$ ions with one 3$d$ electron are usually generated in
slightly reduced \tio. When O is removed, the excess electrons are
unpaired \cite{Oertzen07}. They can occupy the nearby localized Ti
3$d$ orbit and therefore convert Ti$^{4+}$ ions to Ti$^{3+}$ ions.
In a reduced rutile \tio(110) surface, such a defect complex,
Ti$^{3+}$-O$_V$, has been well studied by first-principles
calculations \cite{PhysRevB.55.15919,Oertzen06} and experimentally
by resonant photoelectron diffraction \cite{kruger:055501}.
Therefore, experimental work is needed to clarify whether the
magnetic moments in defective \tio~is due to unpaired 3$d$
electrons localized on Ti$^{3+}$ ions.

Ion irradiation is a non-equilibrium and reproducible method of
inducing defects. Energetic ions displace atoms from their
equilibrium lattice sites, thus creating mainly vacancies and
interstitials. The amount of defects can be controlled by the ion
fluence and energy. In this paper, we irradiated rutile
\tio~single crystals with 2-MeV O ions, resulting in a projected
range of 1.52 $\mu$m and a longitudinal straggling of 0.16 $\mu$m
as calculated by SRIM code (The Stopping and Range of Ions in
Matter) \cite{trim}. As a result of this irradiation, the
formation of Ti/O vacancies/interstitials is expected \cite{trim}.
We selected high-energy oxygen ions as projectiles to avoid the
introduction of foreign elements. Moreover, from a ballistic point
of view, the creation of oxygen vacancies is more efficient, e.g.,
by a factor of 1.5 larger than the Ti-vacancy creation. From SRIM
calculations it is also evident that, at the given energy, the
maximum atomic concentration of the implanted oxygen ions is by a
factor of 500 smaller than the concentration of oxygen recoils.
For the region of maximum defect creation, i.e., at the end of the
ion range, those projectiles play no chemical role. Thus, the
desired local oxygen deficiency can be obtained by the proposed
experimental procedure. The selection of \tio~single crystals
instead of thin films is expected to exclude any effect of
possible magnetic impurities in the underlying substrates
\cite{golmar:262503}, and of the interfaces between films and
substrates \cite{Brinkman}.

Commercial rutile \tio~single crystals were irradiated with 2 MeV
O ions at room temperature with fluences from $1\times10^{15}$ to
$5\times10^{16}$ cm$^{-2}$. All samples are from the same purchase
charge with an equal size of 5$\times$5$\times$0.5 mm$^3$,
corresponding to a mass of 53 mg. The samples thereafter were
named as 1E15, ..., 5E16. All samples were investigated using
SQUID magnetometry (Quantum design MPMS-7XL), X-ray diffraction
(XRD, Siemens D5005), Raman spectroscopy (LabramHR,
Jobin-Yvon-Horiba) and electron spin resonance (ESR, Bruker
ELEXSYS E500 at 9.4 GHz). SQUID measurements indicate that the
virgin \tio~single crystals are weak paramagnets with a
susceptibility of 7.7$\times$10$^{-6}$ emu/T$\cdot$g.

\begin{figure} \center
\includegraphics[scale=0.7]{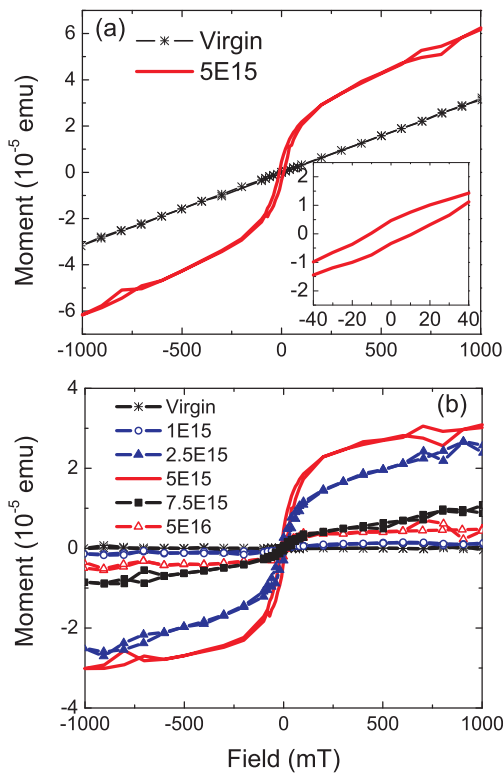}
\caption{(Color online) (a) Magnetic moments measured at 300 K as
a function of magnetic field before and after irradiation. Inset:
the low-field part of the loop for sample 5E15. (b) Magnetic
moments at 300 K for samples with an equal sample-size and
different ion fluences. The linear paramagnetic background has
been subtracted.}\label{fig:squid}
\end{figure}

Fig. \ref{fig:squid}(a) shows the measured magnetic moment of
\tio~at 300 K before and after irradiation with a fluence of
$5\times10^{15}$ cm$^{-2}$. After irradiation the sample shows a
ferromagnetic hysteresis added to the linear paramagnetic
background. The saturation magnetic moment is in the order of
10$^{-5}$ emu, well above the sensitivity limit of SQUID
magnetometry. The inset shows the low-field part in an enlarged
scale. The coercive field of around 10 mT can be clearly resolved.
Fig. \ref{fig:squid}(b) shows the irradiation-fluence dependent
magnetic moment of \tio~single crystals after subtracting the
linear paramagnetic background. The magnetic moment first
increases drastically with increasing fluence, but for the
fluences above $5\times10^{15}$ cm$^{-2}$ it decreases
significantly again. For the 5E15 sample, with the largest
magnetic moment, we also measured the hysteresis loops at
different temperatures (not shown). The loops are weakly
temperature dependent. This is also a feature of the reported
defect-induced ferromagnetism \cite{naaman}.

\begin{figure} \center
\includegraphics[scale=0.57]{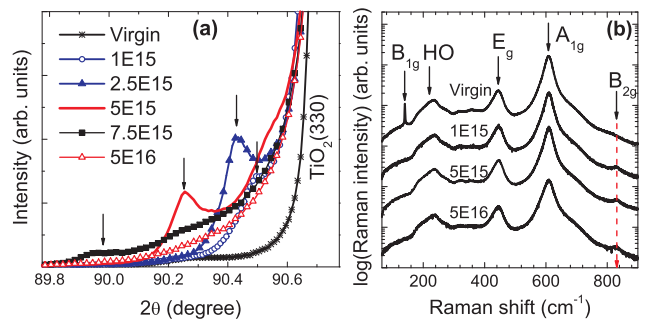}
\caption{(Color online) (a) XRD pattern for O-irradiated \tio. The
peaks (marked by arrows) in the left side of the \tio(330) peak
evidence the irradiation-generated strain in \tio. (b) Raman
spectra of \tio~single crystals after oxygen ion irradiation with
different fluences. The spectra are shifted vertically for better
visualization. A clear mode at the range of \emph{B$_{2g}$} (as
indicated by the dashed arrow) has been observed after
irradiation.}\label{fig:raman}
\end{figure}

The structural evolution upon ion irradiation was assessed by
X-ray diffraction (XRD) and Raman-scattering measurements. Fig.
\ref{fig:raman}(a) shows the XRD patterns close to the \tio(330)
peak before and after irradiation at different fluences. One
remarkable characteristic is that a new peak appears at the left
side of the (330) peak after irradiation. This new peak is induced
by a strained \tio~layer with a larger lattice spacing compared to
that of the virgin \tio. With increasing fluence, this peak shifts
towards smaller 2$\theta$ angles and arrives at a minimum at a
fluence of $7.5\times10^{15}$ cm$^{-2}$. The largest strain is
around 0.50\%, however the peak becomes weaker. With further
increasing fluence, the peak finally disappears and develops into
a gradual shoulder, which means the loss of crystalline long-range
ordering of the irradiated layer. This phenomenon is usually due
to the fact that with increasing density point defects accumulate
and develop into extended defects, e.g., dislocations. The lattice
expansion upon ion irradiation is a common feature in
semiconductors, \eg, in GaN \cite{GaN_expension_review}. 

The effect of oxygen implantation on the structural properties of
\tio~was further studied by Raman spectroscopy. The experiments
were carried out in a micro Raman 180\degr~backscattering
geometry. The scattering intensities were measured using vertical
orientation of the crystal [110]-axis relative to the electric
field vector of the incoming laser. The spectrum of the virgin
sample [Fig. \ref{fig:raman}(b)] shows four distinct lines which
are assigned to the \emph{B$_{1g}$} (143 cm$^{-1}$), \emph{E$_g$}
(447 cm$^{-1}$), and \emph{A$_{1g}$} (612 cm$^{-1}$) fundamental
modes and to a higher order (HO) band around 234 cm{$^{-1}$} of
rutile \tio~\cite{PhysRev.154.522}. The fourth Raman-active
fundamental mode with \emph{B$_{2g}$} symmetry appears as a weak
shoulder at 820 cm$^{-1}$. Neither the fingerprint lines from the
anatase \tio~at 144 cm$^{-1}$ nor those of Ti$_2$O$_3$ at 269
cm$^{-1}$ and at 347 cm{$^{-1}$} have been observed in the
irradiated samples
\cite{Ohsaka78,PhysRevB.3.4253,PhysRevB.7.1131}. Already at the
smallest fluence, the irradiation induces i) the complete
disappearance of \emph{B$_{1g}$} and ii) significant changes in
the range of the \emph{B$_{2g}$} mode, where instead of the
shoulder at 820 cm$^{-1}$ a clearly resolved line appears at 834
cm$^{-1}$ [indicated by the dashed arrows in Fig.
\ref{fig:raman}(b)]. The \emph{B$_{1g}$} mode is very sensitive to
the long-range order of \tio~crystals. It softens when applying
high pressure and looses intensity much faster than the other
fundamental modes \cite{PhysRevB.7.1131}. Raman spectra of micro-
and nanocrystalline \tio~of 5 nm to 2 $\mu$m average crystallite
size did not show the \emph{B$_{1g}$} mode as reported recently
\cite{swamy:163118}. An important conclusion of the complete
absence of \emph{B$_{1g}$} is that the information depth of the
Raman spectra is limited to the irradiated part of the sample. The
integrated intensity of the line at 834 cm$^{-1}$ is $\sim$ 5.5
times larger than that of the corresponding shoulder for a virgin
\tio. In virgin \tio~the \emph{B$_{2g}$} mode represents an
asymmetric (Ti-O) stretching vibration of the (TiO$_6$)-octahedra
(Ti ions are surrounded by six O ions) \cite{swamy:163118}. It is
therefore sensitive to the short-range order and the bond strength
in the crystal, i.e., the local environment of the Ti ions. Since
a reduction of the crystallite size usually causes red-shifts of
vibrational modes, the observed blue-shift by 14 cm$^{-1}$ of the
\emph{B$_{2g}$} mode cannot be explained by phonon confinement.
Furthermore, the line can be neither assigned to an IR-active nor
to a silent mode of rutile \tio~possibily activated due to a
crystal-symmetry reduction \cite{Maroni88}. The line at 834
cm$^{-1}$ indicates a stronger bond between the Ti and the O atoms
than in the virgin rutile. One possible reason is the reduced
coordination number of Ti ions resulting in TiO$_{6-x}$ instead of
TiO$_6$, since a lower coordination number leads to bond
contraction and higher vibrational frequencies. This mechanism
should in principle also affect the other Raman lines,
particularly the \emph{A$_{1g}$} mode. However, a possible
high-energy shoulder would be easily hidden under the large
scattering intensity of the main components. The localized origin
of the mode at 834 cm$^{-1}$ is also supported by the dependence
of the Raman line parameters on implantation fluences. While the
A$_{1g}$ and E$_g$ lines are broadened with increasing oxygen
fluence due to irradiation-induced defects, the local
(TiO$_{6-x}$) stretching mode shows a slight line narrowing.
Therefore, we conclude that oxygen irradiation results in the
local distortion of the overall TiO$_6$ coordination, which gives
rise to a local TiO$_{6-x}$ stretching mode in the range of the
B$_{2g}$ mode. Such a distorted TiO$_{6-x}$ octahedral has also
been considered in a recent experimental and molecular dynamics
simulation study of ion irradiation damage in \tio
\cite{lumpkin:214201}.

\begin{figure} \center
\includegraphics[scale=1]{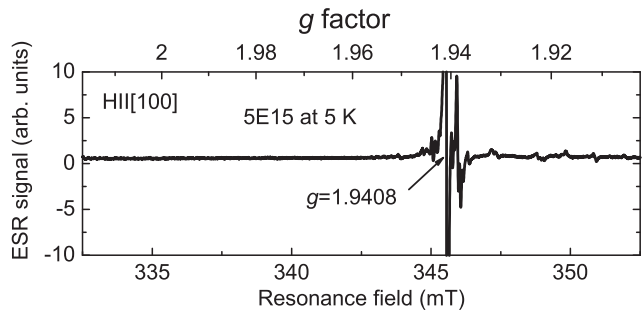}
\caption{ESR signals observed in sample 5E15 at 5 K. The applied
field is along the [100] axis. The main ESR peak accompanied by a
hyperfine interaction pattern has a \emph{g}-factor of 1.9408.
}\label{fig:EPR}
\end{figure}

\begin{figure} \center
\includegraphics[scale=0.75]{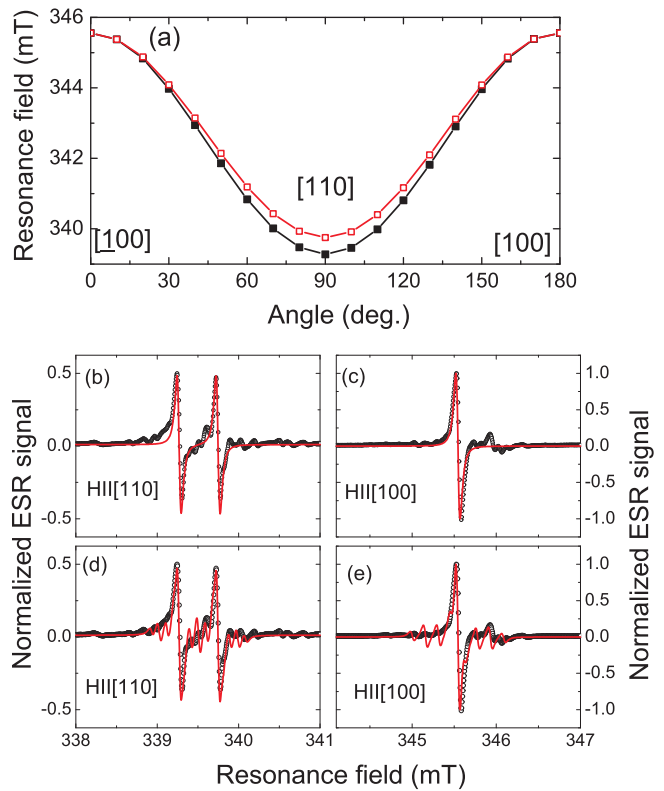}
\caption{(Color online) (a) Angular-dependent resonance field of
the ESR spectrum: it is split into two lines upon field
orientation changing from the [100] to the [110] axis. (b)-(e)
Comparison between simulated (solid lines) and measured (open
circles) ESR data. (b) and (d): With field along the [110]
direction without (b) and with (d) considering hyperfine
interaction. (c) and (e): With field along the [100] direction
without (c) and with (e) considering hyperfine interaction.
}\label{fig:EPR_sim}
\end{figure}

In order to understand the origin of the local moments, ESR
measurements were performed on the virgin and two selected
irradiated samples: 5E15 (with the largest magnetic moment), and
5E16 (with the largest irradiation fluence). ESR directly probes
the interaction between unpaired electrons and the magnetic field.
In a broad applied field range from 50 mT to 550 mT along the
\tio[100] axis and with temperatures down to 5 K, we only detected
one ESR peak at a \emph{g}-factor of 1.9408 accompanied by a
complicated hyperfine interaction pattern for the two irradiated
samples. Fig. \ref{fig:EPR} shows the featured field-range of the
measured ESR spectrum at 5 K for sample 5E15. Note that electrons
trapped by vacancies have a \emph{g}-factor slightly larger than 2
and exhibit no hyperfine features
\cite{sterrer:186101,chong:232502}. The lineshape (see Fig.
\ref{fig:EPR_sim} for details) and the \emph{g}-factor are typical
features in the so-called A spectrum due to Ti$^{3+}$ ions
\cite{chester:2233,PhysRev.135.A144}. Now we come to the question:
how the Ti$^{3+}$ ions are generated after irradiation? Among the
defects created by irradiation, only O vacancies and Ti
interstitials can contribute excess electrons
\cite{Oertzen07,he07}. There are three factors favoring the
formation of Ti$^{3+}$-O$_V$ defect complexes: (1) As the nearest
neighbors O vacancies are more close to substitutional Ti ions,
(2) the number of O vacancies is 1.5 times larger than that of Ti
interstitials \cite{trim} and (3) energetically the formation of O
vacancies is slightly favored over the formation of Ti interstials
\cite{Oertzen07}. Therefore we can conclude that each of
irradiation-induced oxygen vacancies leaves two electrons and
converts the neighboring Ti$^{4+}$ to Ti$^{3+}$ ions, resulting in
Ti$^{3+}$-O$_V$ defect complexes. Such a converting in the reduced
\tio(110) surface has been confirmed recently by both theoretical
and experimental works \cite{Oertzen06,kruger:055501}. It is
important to note that the virgin sample shows no detectable ESR
line down to 5 K (not shown) since the virgin \tio~contains no
unpaired electrons.

The site occupation of the Ti$^{3+}$ ions, namely substitutional
or interstitial, is determined by the angular dependence of the A
spectrum. The crystalline symmetry results in a two-fold splitting
of the Ti$^{3+}$ ESR lines at substitutional sites, and a
four-fold splitting at interstitial sites, when the field
orientation is changed from the [100] to the [110] axis
\cite{PhysRev.135.A144}. As shown in Fig. \ref{fig:EPR_sim}(a), we
can conclude that the Ti$^{3+}$ ions in our samples occupy
substitutional sites. We performed a simulation of the ESR spectra
with this assumption as shown in Fig. \ref{fig:EPR_sim}(b)-(e).
For fields along [110] and [100] and with the anisotropic
\emph{g}-factors [g$_x$, g$_y$, g$_z$]= [1.9408, 1.9738, 1.9766],
we obtained the spectra as shown in Fig. \ref{fig:EPR_sim}(b)-(d).
Without considering the hyperfine interaction, the simulated
spectra agree reasonably well with the experimental data in
\emph{g}-factor, lineshape, and splitting with field rotation.
When taking the hyperfine interaction into account, namely the
nuclear spin of $^{47}$Ti ($I=5/2$) and $^{49}$Ti ($I=7/2$), the
simulations reproduce the experimental results even much better.

The presence of Ti$^{3+}$ at substitutional sites also explains
the XRD results considering the fact that substitutional Ti$^{3+}$
ions are larger than Ti$^{4+}$ ions resulting in the observed
lattice expansion. However, the other point defects, i.e. Ti/O
interstitials can also result in the lattice expansion. Combining
the findings of Raman and ESR measurements, we conclude the
following. Among the various defects created by oxygen ion
irradiation, Ti$^{3+}$ ions on substitutional sites are
accompanied by oxygen vacancies (O$_V$) forming defect complexes,
which result in a local (TiO$_{6-x}$) stretching Raman mode. When
the ion fluence is very large, the irradiated layer loses its
long-range crystalline ordering, such that the fundamental mode of
\tio~is broadened.

Summarizing our results, we conclude that (1) a Ti$^{3+}$-O$_V$
defect complex is generated due to ion irradiation. The Ti$^{3+}$
ions provide local 3$d$ moments which are decisively related with
the observed ferromagnetism; (2) There is an optimum fluence to
produce enough Ti$^{3+}$ ions while keeping the crystalline
ordering. Since ESR is not capable to determine the concentration
of Ti$^{3+}$ ions, it is difficult to get a quantitative insight
of the magnetic coupling mechanism. Despite such an unfolded
question, our research suggests that the combination of ion
irradiation and ESR techniques can be a general approach to verify
the predicted defect-induced ferromagnetism in oxides
\cite{osorio:107203,dev:117204}. The simultaneous observation of
ferromagnetism and ESR signal with a $g$-factor slightly above 2
would be the criterion.

The authors (S.Z. and H.S.) acknowledge financial funding from the
Bundesministerium f\"{u}r Bildung und Forschung (FKZ03N8708).


\begin{thebibliography}{29}
\expandafter\ifx\csname
natexlab\endcsname\relax\def\natexlab#1{#1}\fi
\expandafter\ifx\csname bibnamefont\endcsname\relax
  \def\bibnamefont#1{#1}\fi
\expandafter\ifx\csname bibfnamefont\endcsname\relax
  \def\bibfnamefont#1{#1}\fi
\expandafter\ifx\csname citenamefont\endcsname\relax
  \def\citenamefont#1{#1}\fi
\expandafter\ifx\csname url\endcsname\relax
  \def\url#1{\texttt{#1}}\fi
\expandafter\ifx\csname
urlprefix\endcsname\relax\def\urlprefix{URL }\fi
\providecommand{\bibinfo}[2]{#2}
\providecommand{\eprint}[2][]{\url{#2}}

\bibitem[{\citenamefont{Coey et~al.}(2005)\citenamefont{Coey, Venkatesan,
  Stamenov, Fitzgerald, and Dorneles}}]{coey:024450}
\bibinfo{author}{\bibfnamefont{J.~M.~D.} \bibnamefont{Coey}},
  \bibinfo{author}{\bibfnamefont{M.}~\bibnamefont{Venkatesan}},
  \bibinfo{author}{\bibfnamefont{P.}~\bibnamefont{Stamenov}},
  \bibinfo{author}{\bibfnamefont{C.~B.} \bibnamefont{Fitzgerald}},
  \bibnamefont{and} \bibinfo{author}{\bibfnamefont{L.~S.}
  \bibnamefont{Dorneles}}, \bibinfo{journal}{Phys. Rev. B}
  \textbf{\bibinfo{volume}{72}}, \bibinfo{pages}{024450}
  (\bibinfo{year}{2005}).

\bibitem[{\citenamefont{Yoon et~al.}(2006)\citenamefont{Yoon, Chen, Yang,
  Goodrich, Zuo, Arena, Ziemer, Vittoria, and Harris}}]{yoon06}
\bibinfo{author}{\bibfnamefont{S.~D.} \bibnamefont{Yoon}},
  \bibinfo{author}{\bibfnamefont{Y.}~\bibnamefont{Chen}},
  \bibinfo{author}{\bibfnamefont{A.}~\bibnamefont{Yang}},
  \bibinfo{author}{\bibfnamefont{T.~L.} \bibnamefont{Goodrich}},
  \bibinfo{author}{\bibfnamefont{X.}~\bibnamefont{Zuo}},
  \bibinfo{author}{\bibfnamefont{D.~A.} \bibnamefont{Arena}},
  \bibinfo{author}{\bibfnamefont{K.}~\bibnamefont{Ziemer}},
  \bibinfo{author}{\bibfnamefont{C.}~\bibnamefont{Vittoria}}, \bibnamefont{and}
  \bibinfo{author}{\bibfnamefont{V.~G.} \bibnamefont{Harris}},
  \bibinfo{journal}{J. Phys.: Conden. Matter} \textbf{\bibinfo{volume}{18}},
  \bibinfo{pages}{L355} (\bibinfo{year}{2006}).

\bibitem[{\citenamefont{Hong et~al.}(2006)\citenamefont{Hong, Sakai, Poirot,
  and Briz\'{e}}}]{hong:132404}
\bibinfo{author}{\bibfnamefont{N.~H.} \bibnamefont{Hong}},
  \bibinfo{author}{\bibfnamefont{J.}~\bibnamefont{Sakai}},
  \bibinfo{author}{\bibfnamefont{N.}~\bibnamefont{Poirot}}, \bibnamefont{and}
  \bibinfo{author}{\bibfnamefont{V.}~\bibnamefont{Briz\'{e}}},
  \bibinfo{journal}{Phys. Rev. B} \textbf{\bibinfo{volume}{73}},
  \bibinfo{pages}{132404} (\bibinfo{year}{2006}).

\bibitem[{\citenamefont{Zhou et~al.}(2008)\citenamefont{Zhou, Potzger, Talut,
  Reuther, Kuepper, Grenzer, Xu, M\"{u}cklich, Helm, Fassbender
  et~al.}}]{defects_zhou}
\bibinfo{author}{\bibfnamefont{S.}~\bibnamefont{Zhou}},
  \bibinfo{author}{\bibfnamefont{K.}~\bibnamefont{Potzger}},
  \bibinfo{author}{\bibfnamefont{G.}~\bibnamefont{Talut}},
  \bibinfo{author}{\bibfnamefont{H.}~\bibnamefont{Reuther}},
  \bibinfo{author}{\bibfnamefont{K.}~\bibnamefont{Kuepper}},
  \bibinfo{author}{\bibfnamefont{J.}~\bibnamefont{Grenzer}},
  \bibinfo{author}{\bibfnamefont{Q.}~\bibnamefont{Xu}},
  \bibinfo{author}{\bibfnamefont{A.}~\bibnamefont{M\"{u}cklich}},
  \bibinfo{author}{\bibfnamefont{M.}~\bibnamefont{Helm}},
  \bibinfo{author}{\bibfnamefont{J.}~\bibnamefont{Fassbender}},
  \bibinfo{author}{\bibfnamefont{E.}~\bibnamefont{Arenholz}},
  \bibinfo{journal}{J. Phys. D: Appl. Phys.}
  \textbf{\bibinfo{volume}{41}}, \bibinfo{pages}{105011}
  (\bibinfo{year}{2008}).

\bibitem[{\citenamefont{Elfimov et~al.}(2002)\citenamefont{Elfimov, Yunoki, and
  Sawatzky}}]{elfimov02}
\bibinfo{author}{\bibfnamefont{I.~S.} \bibnamefont{Elfimov}},
  \bibinfo{author}{\bibfnamefont{S.}~\bibnamefont{Yunoki}}, \bibnamefont{and}
  \bibinfo{author}{\bibfnamefont{G.~A.} \bibnamefont{Sawatzky}},
  \bibinfo{journal}{Phys. Rev. Lett.} \textbf{\bibinfo{volume}{89}},
  \bibinfo{pages}{216403} (\bibinfo{year}{2002}).

\bibitem[{\citenamefont{Osorio-Guill\'{e}n
  et~al.}(2006)\citenamefont{Osorio-Guill\'{e}n, Lany, Barabash, and
  Zunger}}]{osorio:107203}
\bibinfo{author}{\bibfnamefont{J.}~\bibnamefont{Osorio-Guill\'{e}n}},
  \bibinfo{author}{\bibfnamefont{S.}~\bibnamefont{Lany}},
  \bibinfo{author}{\bibfnamefont{S.~V.} \bibnamefont{Barabash}},
  \bibnamefont{and} \bibinfo{author}{\bibfnamefont{A.}~\bibnamefont{Zunger}},
  \bibinfo{journal}{Phys. Rev. Lett.} \textbf{\bibinfo{volume}{96}},
  \bibinfo{pages}{107203} (\bibinfo{year}{2006}).

\bibitem[{\citenamefont{Chanier et~al.}(2008)\citenamefont{Chanier, Opahle,
  Sargolzaei, Hayn, and Lannoo}}]{chanier:026405}
\bibinfo{author}{\bibfnamefont{T.}~\bibnamefont{Chanier}},
  \bibinfo{author}{\bibfnamefont{I.}~\bibnamefont{Opahle}},
  \bibinfo{author}{\bibfnamefont{M.}~\bibnamefont{Sargolzaei}},
  \bibinfo{author}{\bibfnamefont{R.}~\bibnamefont{Hayn}}, \bibnamefont{and}
  \bibinfo{author}{\bibfnamefont{M.}~\bibnamefont{Lannoo}},
  \bibinfo{journal}{Phys. Rev. Lett.} \textbf{\bibinfo{volume}{100}},
  \bibinfo{pages}{026405} (\bibinfo{year}{2008}).

\bibitem[{\citenamefont{Oertzen and Gerson}(2007)}]{Oertzen07}
\bibinfo{author}{\bibfnamefont{G.~U.} \bibnamefont{Oertzen}} \bibnamefont{and}
  \bibinfo{author}{\bibfnamefont{A.~R.} \bibnamefont{Gerson}},
  \bibinfo{journal}{J. Phys. Chem. Solids} \textbf{\bibinfo{volume}{68}},
  \bibinfo{pages}{324} (\bibinfo{year}{2007}).

\bibitem[{\citenamefont{Lindan et~al.}(1997)\citenamefont{Lindan, Harrison,
  Gillan, and White}}]{PhysRevB.55.15919}
\bibinfo{author}{\bibfnamefont{P.~J.~D.} \bibnamefont{Lindan}},
  \bibinfo{author}{\bibfnamefont{N.~M.} \bibnamefont{Harrison}},
  \bibinfo{author}{\bibfnamefont{M.~J.} \bibnamefont{Gillan}},
  \bibnamefont{and} \bibinfo{author}{\bibfnamefont{J.~A.} \bibnamefont{White}},
  \bibinfo{journal}{Phys. Rev. B} \textbf{\bibinfo{volume}{55}},
  \bibinfo{pages}{15919} (\bibinfo{year}{1997}).

\bibitem[{\citenamefont{Oertzen and Gerson}(2006)}]{Oertzen06}
\bibinfo{author}{\bibfnamefont{G.~U.} \bibnamefont{Oertzen}} \bibnamefont{and}
  \bibinfo{author}{\bibfnamefont{A.~R.} \bibnamefont{Gerson}},
  \bibinfo{journal}{Int. J. Quantum Chem} \textbf{\bibinfo{volume}{106}},
  \bibinfo{pages}{2054} (\bibinfo{year}{2006}).

\bibitem[{\citenamefont{Kr\"{u}ger et~al.}(2008)\citenamefont{Kr\"{u}ger,
  Bourgeois, Domenichini, Magnan, Chandesris, F\`{e}vre, Flank, Jupille,
  Floreano, Cossaro et~al.}}]{kruger:055501}
\bibinfo{author}{\bibfnamefont{P.}~\bibnamefont{Kr\"{u}ger}},
  \bibinfo{author}{\bibfnamefont{S.}~\bibnamefont{Bourgeois}},
  \bibinfo{author}{\bibfnamefont{B.}~\bibnamefont{Domenichini}},
  \bibinfo{author}{\bibfnamefont{H.}~\bibnamefont{Magnan}},
  \bibinfo{author}{\bibfnamefont{D.}~\bibnamefont{Chandesris}},
  \bibinfo{author}{\bibfnamefont{P.~L.} \bibnamefont{F\`{e}vre}},
  \bibinfo{author}{\bibfnamefont{A.~M.} \bibnamefont{Flank}},
  \bibinfo{author}{\bibfnamefont{J.}~\bibnamefont{Jupille}},
  \bibinfo{author}{\bibfnamefont{L.}~\bibnamefont{Floreano}},
  \bibinfo{author}{\bibfnamefont{A.}~\bibnamefont{Cossaro}},
  \bibinfo{author}{\bibfnamefont{A.}~\bibnamefont{Verdini}},
  \bibinfo{author}{\bibfnamefont{A.}~\bibnamefont{Morgante}},
  \bibinfo{journal}{Phys. Rev. Lett.}
  \textbf{\bibinfo{volume}{100}}, \bibinfo{eid}{055501} (\bibinfo{year}{2008}).

\bibitem[{\citenamefont{Ziegler et~al.}(1985)\citenamefont{Ziegler, Biersack,
  and Littmark}}]{trim}
\bibinfo{author}{\bibfnamefont{J.}~\bibnamefont{Ziegler}},
  \bibinfo{author}{\bibfnamefont{J.}~\bibnamefont{Biersack}}, \bibnamefont{and}
  \bibinfo{author}{\bibfnamefont{U.}~\bibnamefont{Littmark}},
  \emph{\bibinfo{title}{The stopping and range of ions in matter}}
  (\bibinfo{publisher}{Pergamon}, \bibinfo{address}{New York},
  \bibinfo{year}{1985}).

\bibitem[{\citenamefont{Golmar et~al.}(2008)\citenamefont{Golmar, Navarro,
  Torres, S\'{a}nchez, Saccone, dos Santos~Claro, Ben\'{\i}tez, and
  Schilardi}}]{golmar:262503}
\bibinfo{author}{\bibfnamefont{F.}~\bibnamefont{Golmar}},
  \bibinfo{author}{\bibfnamefont{A.~M.~M.} \bibnamefont{Navarro}},
  \bibinfo{author}{\bibfnamefont{C.~E.~R.} \bibnamefont{Torres}},
  \bibinfo{author}{\bibfnamefont{F.~H.} \bibnamefont{S\'{a}nchez}},
  \bibinfo{author}{\bibfnamefont{F.~D.} \bibnamefont{Saccone}},
  \bibinfo{author}{\bibfnamefont{P.~C.} \bibnamefont{dos Santos~Claro}},
  \bibinfo{author}{\bibfnamefont{G.~A.} \bibnamefont{Ben\'{\i}tez}},
  \bibnamefont{and} \bibinfo{author}{\bibfnamefont{P.~L.}
  \bibnamefont{Schilardi}}, \bibinfo{journal}{Appl. Phys. Lett.}
  \textbf{\bibinfo{volume}{92}}, \bibinfo{pages}{262503}
  (\bibinfo{year}{2008}).

\bibitem[{\citenamefont{Brinkman et~al.}(2007)\citenamefont{Brinkman, Huijben,
  Van~Zalk, Huijben, Zeitler, Maan, Van~der Wiel, Rijnders, Blank, and
  Hilgenkamp}}]{Brinkman}
\bibinfo{author}{\bibfnamefont{A.}~\bibnamefont{Brinkman}},
  \bibinfo{author}{\bibfnamefont{M.}~\bibnamefont{Huijben}},
  \bibinfo{author}{\bibfnamefont{M.}~\bibnamefont{Van~Zalk}},
  \bibinfo{author}{\bibfnamefont{J.}~\bibnamefont{Huijben}},
  \bibinfo{author}{\bibfnamefont{U.}~\bibnamefont{Zeitler}},
  \bibinfo{author}{\bibfnamefont{J.~C.} \bibnamefont{Maan}},
  \bibinfo{author}{\bibfnamefont{W.~G.} \bibnamefont{Van~der Wiel}},
  \bibinfo{author}{\bibfnamefont{G.}~\bibnamefont{Rijnders}},
  \bibinfo{author}{\bibfnamefont{D.~H.~A.} \bibnamefont{Blank}},
  \bibnamefont{and}
  \bibinfo{author}{\bibfnamefont{H.}~\bibnamefont{Hilgenkamp}},
  \bibinfo{journal}{Nat. Mater.} \textbf{\bibinfo{volume}{6}},
  \bibinfo{pages}{493} (\bibinfo{year}{2007}).

\bibitem[{\citenamefont{Kopnov et~al.}(2007)\citenamefont{Kopnov, Vager, and
  Naaman}}]{naaman}
\bibinfo{author}{\bibfnamefont{G.}~\bibnamefont{Kopnov}},
  \bibinfo{author}{\bibfnamefont{Z.}~\bibnamefont{Vager}}, \bibnamefont{and}
  \bibinfo{author}{\bibfnamefont{R.}~\bibnamefont{Naaman}},
  \bibinfo{journal}{Adv. Mater.} \textbf{\bibinfo{volume}{19}},
  \bibinfo{pages}{925} (\bibinfo{year}{2007}).

\bibitem[{\citenamefont{Ronning et~al.}(2001)\citenamefont{Ronning, Carlsonb,
  and Davis}}]{GaN_expension_review}
\bibinfo{author}{\bibfnamefont{C.}~\bibnamefont{Ronning}},
  \bibinfo{author}{\bibfnamefont{E.~P.} \bibnamefont{Carlsonb}},
  \bibnamefont{and} \bibinfo{author}{\bibfnamefont{R.~F.} \bibnamefont{Davis}},
  \bibinfo{journal}{Phys. Rep.} \textbf{\bibinfo{volume}{351}},
  \bibinfo{pages}{349} (\bibinfo{year}{2001}).

\bibitem[{\citenamefont{Porto et~al.}(1967)\citenamefont{Porto, Fleury, and
  Damen}}]{PhysRev.154.522}
\bibinfo{author}{\bibfnamefont{S.~P.~S.} \bibnamefont{Porto}},
  \bibinfo{author}{\bibfnamefont{P.~A.} \bibnamefont{Fleury}},
  \bibnamefont{and} \bibinfo{author}{\bibfnamefont{T.~C.} \bibnamefont{Damen}},
  \bibinfo{journal}{Phys. Rev.} \textbf{\bibinfo{volume}{154}},
  \bibinfo{pages}{522} (\bibinfo{year}{1967}).

\bibitem[{\citenamefont{Ohsaka et~al.}(1978)\citenamefont{Ohsaka, Izumi, and
  Fujiki}}]{Ohsaka78}
\bibinfo{author}{\bibfnamefont{T.}~\bibnamefont{Ohsaka}},
  \bibinfo{author}{\bibfnamefont{F.}~\bibnamefont{Izumi}}, \bibnamefont{and}
  \bibinfo{author}{\bibfnamefont{Y.}~\bibnamefont{Fujiki}},
  \bibinfo{journal}{J. Raman Spectrosc.} \textbf{\bibinfo{volume}{7}},
  \bibinfo{pages}{321} (\bibinfo{year}{1978}).

\bibitem[{\citenamefont{Mooradian and Raccah}(1971)}]{PhysRevB.3.4253}
\bibinfo{author}{\bibfnamefont{A.}~\bibnamefont{Mooradian}} \bibnamefont{and}
  \bibinfo{author}{\bibfnamefont{P.~M.} \bibnamefont{Raccah}},
  \bibinfo{journal}{Phys. Rev. B} \textbf{\bibinfo{volume}{3}},
  \bibinfo{pages}{4253} (\bibinfo{year}{1971}).

\bibitem[{\citenamefont{Samara and Peercy}(1973)}]{PhysRevB.7.1131}
\bibinfo{author}{\bibfnamefont{G.~A.} \bibnamefont{Samara}} \bibnamefont{and}
  \bibinfo{author}{\bibfnamefont{P.~S.} \bibnamefont{Peercy}},
  \bibinfo{journal}{Phys. Rev. B} \textbf{\bibinfo{volume}{7}},
  \bibinfo{pages}{1131} (\bibinfo{year}{1973}).

\bibitem[{\citenamefont{Swamy et~al.}(2006)\citenamefont{Swamy, Muddle, and
  Dai}}]{swamy:163118}
\bibinfo{author}{\bibfnamefont{V.}~\bibnamefont{Swamy}},
  \bibinfo{author}{\bibfnamefont{B.~C.} \bibnamefont{Muddle}},
  \bibnamefont{and} \bibinfo{author}{\bibfnamefont{Q.}~\bibnamefont{Dai}},
  \bibinfo{journal}{Appl. Phys. Lett.} \textbf{\bibinfo{volume}{89}},
  \bibinfo{pages}{163118} (\bibinfo{year}{2006}).

\bibitem[{\citenamefont{Maroni}(1988)}]{Maroni88}
\bibinfo{author}{\bibfnamefont{V.}~\bibnamefont{Maroni}}, \bibinfo{journal}{J.
  Phys. Chem. Sol.} \textbf{\bibinfo{volume}{49}}, \bibinfo{pages}{307}
  (\bibinfo{year}{1988}).

\bibitem[{\citenamefont{Lumpkin et~al.}(2008)\citenamefont{Lumpkin, Smith,
  Blackford, Thomas, Whittle, Marks, and Zaluzec}}]{lumpkin:214201}
\bibinfo{author}{\bibfnamefont{G.~R.} \bibnamefont{Lumpkin}},
  \bibinfo{author}{\bibfnamefont{K.~L.} \bibnamefont{Smith}},
  \bibinfo{author}{\bibfnamefont{M.~G.} \bibnamefont{Blackford}},
  \bibinfo{author}{\bibfnamefont{B.~S.} \bibnamefont{Thomas}},
  \bibinfo{author}{\bibfnamefont{K.~R.} \bibnamefont{Whittle}},
  \bibinfo{author}{\bibfnamefont{N.~A.} \bibnamefont{Marks}}, \bibnamefont{and}
  \bibinfo{author}{\bibfnamefont{N.~J.} \bibnamefont{Zaluzec}},
  \bibinfo{journal}{Phys. Rev. B} \textbf{\bibinfo{volume}{77}},
  \bibinfo{pages}{214201} (\bibinfo{year}{2008}).

\bibitem[{\citenamefont{Sterrer et~al.}(2005)\citenamefont{Sterrer, Fischbach,
  Risse, and Freund}}]{sterrer:186101}
\bibinfo{author}{\bibfnamefont{M.}~\bibnamefont{Sterrer}},
  \bibinfo{author}{\bibfnamefont{E.}~\bibnamefont{Fischbach}},
  \bibinfo{author}{\bibfnamefont{T.}~\bibnamefont{Risse}}, \bibnamefont{and}
  \bibinfo{author}{\bibfnamefont{H.-J.} \bibnamefont{Freund}},
  \bibinfo{journal}{Phys. Rev. Lett.} \textbf{\bibinfo{volume}{94}},
  \bibinfo{pages}{186101} (\bibinfo{year}{2005}).

\bibitem[{\citenamefont{Chong et~al.}(2008)\citenamefont{Chong, Kadowaki, Xia,
  and Idriss}}]{chong:232502}
\bibinfo{author}{\bibfnamefont{S.~V.} \bibnamefont{Chong}},
  \bibinfo{author}{\bibfnamefont{K.}~\bibnamefont{Kadowaki}},
  \bibinfo{author}{\bibfnamefont{J.}~\bibnamefont{Xia}}, \bibnamefont{and}
  \bibinfo{author}{\bibfnamefont{H.}~\bibnamefont{Idriss}},
  \bibinfo{journal}{Appl. Phys. Lett.} \textbf{\bibinfo{volume}{92}},
  \bibinfo{pages}{232502} (\bibinfo{year}{2008}).

\bibitem[{\citenamefont{Chester}(1961)}]{chester:2233}
\bibinfo{author}{\bibfnamefont{P.~F.} \bibnamefont{Chester}},
  \bibinfo{journal}{J. Appl. Phys.} \textbf{\bibinfo{volume}{32}},
  \bibinfo{pages}{2233} (\bibinfo{year}{1961}).

\bibitem[{\citenamefont{Yamaka and Barnes}(1964)}]{PhysRev.135.A144}
\bibinfo{author}{\bibfnamefont{E.}~\bibnamefont{Yamaka}} \bibnamefont{and}
  \bibinfo{author}{\bibfnamefont{R.~G.} \bibnamefont{Barnes}},
  \bibinfo{journal}{Phys. Rev.} \textbf{\bibinfo{volume}{135}},
  \bibinfo{pages}{A144} (\bibinfo{year}{1964}).

\bibitem[{\citenamefont{He et~al.}(2007)\citenamefont{He, Behera, Finnis, Li,
  Dickey, Phillpot, and Sinnott}}]{he07}
\bibinfo{author}{\bibfnamefont{J.}~\bibnamefont{He}},
  \bibinfo{author}{\bibfnamefont{R.~K.} \bibnamefont{Behera}},
  \bibinfo{author}{\bibfnamefont{M.~W.} \bibnamefont{Finnis}},
  \bibinfo{author}{\bibfnamefont{X.}~\bibnamefont{Li}},
  \bibinfo{author}{\bibfnamefont{E.~C.} \bibnamefont{Dickey}},
  \bibinfo{author}{\bibfnamefont{S.~R.} \bibnamefont{Phillpot}},
  \bibnamefont{and} \bibinfo{author}{\bibfnamefont{S.~B.}
  \bibnamefont{Sinnott}}, \bibinfo{journal}{Acta Mater.}
  \textbf{\bibinfo{volume}{55}}, \bibinfo{pages}{4325} (\bibinfo{year}{2007}).

\bibitem[{\citenamefont{Dev et~al.}(2008)\citenamefont{Dev, Xue, and
  Zhang}}]{dev:117204}
\bibinfo{author}{\bibfnamefont{P.}~\bibnamefont{Dev}},
  \bibinfo{author}{\bibfnamefont{Y.}~\bibnamefont{Xue}}, \bibnamefont{and}
  \bibinfo{author}{\bibfnamefont{P.}~\bibnamefont{Zhang}},
  \bibinfo{journal}{Phys. Rev. Lett.} \textbf{\bibinfo{volume}{100}},
  \bibinfo{pages}{117204} (\bibinfo{year}{2008}).

\end{thebibliography}

\end{document}